\documentclass[pra,showpacs,twocolumn]{revtex4}

\usepackage{hyperref,fontenc,times,amsmath,amsfonts,amssymb,graphics,graphicx,epsfig,color,bbm,bm,units}

\newcommand{\dd}{\mathrm{d}}
\newcommand{\ee}{\mathrm{e}}
\newcommand{\ii}{\mathrm{i}}

\begin{document}

\title{Estimation of Phase and Diffusion: Combining  Quantum Statistics and Classical Noise}
\author{Sergey I.\ Knysh}\email{Sergey.I.Knysh@nasa.gov}

\author{Gabriel A.\ Durkin} \email{Gabriel.Durkin@nasa.gov}
\affiliation{Quantum Artificial Intelligence Laboratory (QuAIL), NASA Ames Research Center,  Moffett Field, California 94035, USA}
\date{\today}
\begin{abstract}
Coherent ensembles of $N$ qubits present an advantage in quantum phase estimation over separable mixtures, but coherence decay due to classical phase diffusion reduces overall precision. In some contexts, the strength of diffusion may be the parameter of interest. We examine estimation of both phase and diffusion in large spin systems using a novel mathematical formulation.  For the first time, we show a closed form expression for the quantum Fisher information for estimation of a unitary parameter in a noisy environment. The optimal probe state has a non-Gaussian profile and differs also from the canonical phase state; it saturates a new tight precision bound.   For noise below a critical threshold, entanglement always leads to enhanced precision, but the shot-noise limit is beaten only by a constant factor, independent of $N$. We provide upper and lower bounds to this factor, valid in low and high noise regimes. Unlike other noise types, it is shown for $N \gg 1$  that phase and diffusion can be measured simultaneously and optimally by canonical phase measurements. 
\end{abstract}
\pacs{42.50.-p,42.50.St,06.20.Dk}

\maketitle
Dephasing, or a random uncontrollable phase accumulation, is one of the most important types of noise in quantum systems, responsible for a transition from quantum to classical behaviour. It is a dissipationless noise; no energy or particles disappear from the system. It has relevance for metrology with atom and spin ensembles, where the particle number is conserved \cite{firstpaper,dephasing10}. It also plays a role in optic-fiber interferometric sensors \cite{fibershapesensor} where thermal perturbations and mechanical strains can lead to measurable diffusion in both interferometric phase and polarisation of light. Shape sensors woven from fiber arrays embedded in aircraft wings subjected to turbulent airflow provide precursors to structural failure \cite{fibershapesensor2}, as could similar sensors placed on instrument surfaces  of deep-space telescopes exposed to solar heating and vibration  \cite{spacetelescope}. In this paper, we explore quantum estimation of both phase and collective dephasing (or drift and diffusion parameters) as a step towards revealing advantages offered by quantum instruments and sensors in scenarios such as these. This is a departure from much previous work, which examined local or intrinsic diffusion, as occurs when each qubit or atom is subject to its own independent dephasing mechanism  \cite{firstpaper,RDD12,diffother}. (We shall see later that collective dephasing has a stronger effect in reducing quantum coherences than local dephasing.) A general overview of the field of quantum metrology is provided in Refs. \cite{Giovanetti04,Giovanetti11}. 

For systems of small particle number $N$, finding optimal quantum states and precision bounds can be approached numerically \cite{OxfordNum}, but this becomes intractable for increasing $N$. Should quantum correlations offer favorable scaling of measurement error with $N$, then the limit $N \gg 1$ is the interesting and relevant one, where the greatest benefits lie. Here, as in \cite{knysh}, we focus  on calculations in the asymptotic limit $N \sim \infty$; yet in comparison with numerical data for dephasing it emerges that convergence to leading  asymptotic behaviour is already established for modest ensembles of  $10$ to $100$ particles. Numerical optimization results for $N=80$ are shown in FIG.1, and compared to analytically derived expressions. The spin formalism with total spin $j=N/2$ we employ has some universality in its scope;  for $N=1$ it can represent a superconducting flux qubit \cite{FluxMZ}, or for $N \gg1$ an ensemble of atoms in a double-well potential \cite{BECMZ},   and any two-mode interferometer via the Schwinger isomorphism \cite{Yurke}.

\begin{figure}
\includegraphics[width=3.5in]{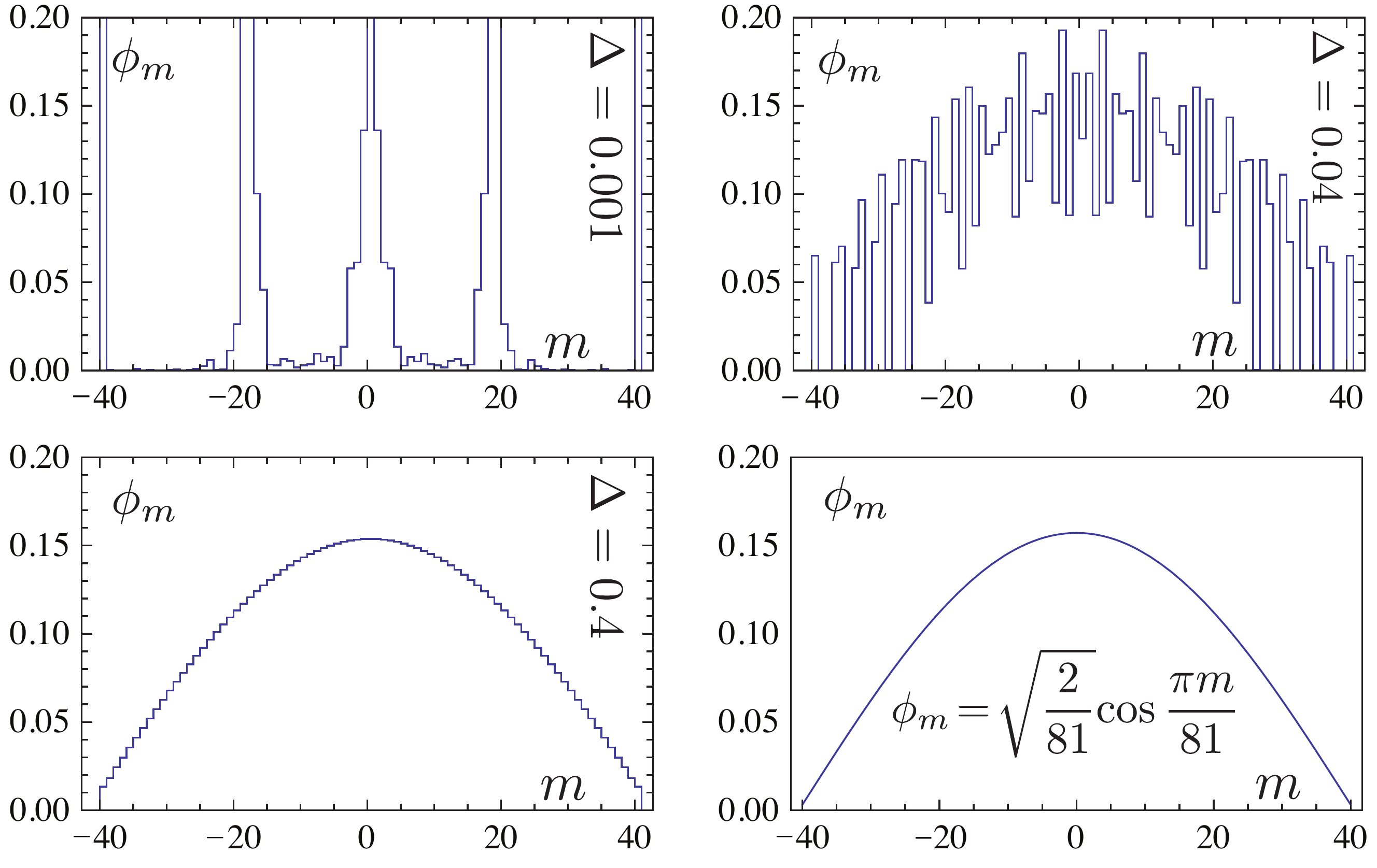}
\caption{\label{bars}
\small{The discrete amplitudes $\phi_m$ of the optimal state found numerically for various strengths of dephasing $\Delta \in \{0.001, 0.04, 0.4\}$ or effective masses $M= \Delta j^2 \in \{1.6,64,640\}$ for a system of spin $j=40$, equivalent to $80$ photons distributed between two interferometric modes. The transition from a series of delta spikes to a smooth unimodal distribution is apparent. For very small dephasing there is still residual NOON state contribution to the optimal state, indicated by  large components at $m= \pm j$. The (smooth) asymptotically optimal Cosine state of eqn.\eqref{optimal} is at bottom right. }}
\end{figure}

Phase precision for a single qubit (N=1) under dephasing has been solved exactly in \cite{qubitresult}. Recently, Genoni et al. presented numerical and experimental work examining the structure of optimal Gaussian states, i.e. families of squeezed, thermal and coherent states for phase estimation in the presence of collective dephasing \cite{Paris12,Paris11}.  By exploiting a novel purification scheme  \cite{Escher12},  upper bounds on phase precision under collective dephasing may be found -- though it was not known til now whether these bounds were tight. Neither was it known which optimal states could approach these bounds. The formulation of tight precision bounds and optimal states for the estimation of the dephasing strength itself has not been addressed at all. 

\emph{Dynamics}: Before we approach phase estimation, let us first examine the diffusion process. The quantum master equation governing both unitary phase evolution and decoherence via phase diffusion is very simple, with quantum spin operator $J_z$ responsible for both processes. An ensemble of $N$ qubits, spins or polarized photons is represented by a density matrix $\rho$ of $2j+1 = N+1$ dimensions, spanned by orthonormal eigenstates $\{|m\rangle\}$ of $J_z$, where $J_z |m \rangle = m | m \rangle $ and $m \in \{-j, -j+1, ... ,+j\}$. The phase/dephasing master equation is \cite{mastereqn}:
\begin{equation}
\frac{d \rho}{d \theta} = -i [J_z, \rho ]  - \frac{\gamma}{2}  [J_z,  [J_z, \rho]   ]   \end{equation} with $\theta$ a time-like variable and operator commutator $[A,B] = A B - B A$. 
The first commutator on the right side gives rise to the unitary drift dynamics, and the double commutator leads to phase diffusion or dephasing.
Similar dynamics have been discussed recently in quantum control of phase diffusion within Josephson junctions \cite{JJdiffusion}. The non-unitary dynamics for $\gamma >0$ arises in two-mode Bose-Einstein condensates due to collisions \cite{dephasing10}, or alternatively, due to back-action of an external optical field \cite{phD}. (This master equation also describes photon dynamics in an interferometer, with dephasing a consequence of the radiation pressure on one of the mirrors \cite{Escher12}.) The `dephased'  state has density matrix elements as follows:
\begin{equation}
  \rho_{m m'}^{\left( \theta \right)} = \ee^{- \frac{\Delta}{2} \left( m -
  m' \right)^2 - \ii \left( m - m' \right) \theta} \phi_m^{\ast} \phi_{m'},
  \label{rhom}
\end{equation}
where $\left| \phi \right\rangle = \sum_{m = - j}^j \phi_m \left| m
\right\rangle$ is the initial state, $\theta$ is the phase to be estimated and $\Delta= \int \gamma (d \theta)$ is the dephasing
parameter. To simplify calculation we restrict amplitudes $\left\{ \phi_m
\right\}$ to real values, which is the optimal choice. The small dephasing case $\Delta \ll 1$ is the interesting limit, in contrast to
$\Delta \gtrsim 1$, when off-diagonal matrix elements become completely suppressed (producing a state that is increasingly symmetric under any phase evolution and useless as a `pointer'). We will show that above a critical $\Delta_c \sim 0.25$ any ensemble of $N$ spins should be applied in series, one at a time, to the phase estimation task. Thus we expect the small $\Delta \ll \Delta_c$ regime is where large-scale entangled states will be useful.

\emph{Asymptotic Limit}: Ultimate precision in parameter estimation is quantified by Quantum Fisher Information (QFI) \cite{Braunstein94,ParisQmTech}, although other metrics exist  \cite{HowardPRX,HuelgaPRL}. QFI is a function of the initial quantum probe state and the dynamics to which it is subjected; both those dynamics that encode the parameter, and those due to noise. For a single parameter such as $\theta$ the reciprocal of QFI provides a lower bound to mean squared error that is saturable for large data sets. It is straightforward to compute for pure state $| \psi \rangle$ evolved by $\exp - i \theta J_z$  for $\gamma =0$ above; it is equivalent to $4 ( \langle \psi | J_z^2 | \psi \rangle  - \langle \psi | J_z | \psi \rangle^2  )$. 

For mixed states, as will occur under noisy dynamics, the QFI 
requires diagonalization of the density matrix. This is not an easy task in general,
though it is made analytically tractable by considering the asymptotic limit $j \gg 1$. In this
limit we approximate the discrete spin projection `$m$' index by a continuous variable $\frac{m}{j} \equiv x \in \left[ - 1 ; 1
\right]$:
\begin{equation}
  \label{rhox}
  \begin{array}{ll}
    \phi_m \rightarrow \frac{1}{\sqrt{j}} \phi \left( x \right), & \rho_{m
    m'}^{\left( 0 \right)} \rightarrow \ee^{- \frac{j^2 \Delta}{2} \left( x
    - y \right)^2} \phi \left( x \right) \phi \left( y \right),
  \end{array}
  \tag{\ref{rhom}$'$}
\end{equation}
where we set $\theta = 0$ to make the density real-valued. (Quantum Fisher information cannot depend on the particular value of the unitary parameter or phase \cite{ParisQmTech}.) The interval of valid values of $x$
can be extended to $x \in \left( - \infty, \infty \right)$. We need not mind
that the values of $x$ are bounded as long as we remember to impose a boundary
condition $\phi \left( x \right) = 0$ for $\left| x \right| > 1$.

Next we recognize the Gaussian kernel in eqn.(\ref{rhox}) as the free-particle
Green's function to rewrite the density matrix in a representation-free operator form:
\begin{equation}
  \rho = \ee^{- U / 2} \ee^{- T} \ee^{- U / 2}, \label{eee}
\end{equation}
with `potential' $U = - \ln \phi^2 \left( x \right)$, an operator diagonal
in $x$-representation; and the `kinetic energy' \ $T = \frac{1}{2} \ln
\frac{2 \pi}{M} + \frac{P^2}{2 M}$, with `momentum' operator  $P = -
\ii \frac{\partial}{\partial x}$ . `Mass' $M = j^2 \Delta$ will serve
as a large parameter in the expansion.

Using a Baker-Campbell-Hausdorff identity, we rewrite eqn.(\ref{eee}):
\begin{equation}
  \rho = \ee^{- H} = \ee^{- T - U - \frac{1}{12} \left[ T, \left[ T, U
  \right] \right] - \frac{1}{24} \left[ U, \left[ T, U \right] \right] -
  \cdots} \label{BCH}
\end{equation}
To leading order, $H_0 = T + U$, corresponds to a simple quantum mechanics
problem; higher order commutators represent subsequent orders of a WKB-like
expansion, although the symmetrically-split operators in eqn.\eqref{eee}  produce no second-order term.  The
necessary condition for such an expansion in inverse powers of $M$ to converge
is that the potential $U \left( x \right)$, i.e. the wavefunction $\phi \left(
x \right)$, is smooth. The higher-order terms are essential to recovering the correct large $N$ functionality of the Fisher information, as discussed in the next section (more details in the Appendix).

\emph{Optimizing Phase Precision}: Quantum Fisher information may be written generally as $F_\theta = j^2 \text{Tr} [\rho L_{\theta}^2]$, expressed in terms of the symmetric logarithmic
derivative $jL_\theta$ (factor of $j$ incorporated for convenience) that solves
\begin{equation}
  \frac{1}{2} \left\{ \rho, jL_{\theta} \right\} =  - \ii \left[ jX, \rho \right],
  \label{defSLD}
\end{equation}
where $X = J_z / j$ is the new `coordinate' operator, and the operator anti-commutator $\{A,B\} =A B - B A$. The calculation of $L_\theta$ and then $F_\theta$ is presented in Appendix \ref{append}. A  novel  and universal result for $L_\theta$ (it  holds generally for the symmetric logarithmic derivative of any unitary shift) is its formulation as a series of nested commutators, following the Taylor expansion of the hyperbolic tangent:
\begin{equation}
  L_{\theta} \! =  \!- 2 \ii \tanh  (  \left[ H, \bullet \right] /2) X , \label{SLD}
\end{equation}
where $ \left[ H, \bullet \right] O =  \left[ H, O \right] $ and $ \left[ H, \bullet \right] ^2 O =  \left[ H, \left[ H, O \right] \right] $, etc. Utilizing this result, the first non-trivial contribution to Fisher information is $\langle U'' \rangle$ (primes denote derivatives with respect to $x$). This $U''$ represents an `information potential' in the Bohm formulation of quantum mechanics (previously linked with Fisher information in Refs.\cite{BohmQmPot}).
We find the overall result simplifies to 
\begin{equation}
  F_{\theta} = \frac{1}{\Delta} - \frac{1}{M \Delta} \int \phi'^{ \: 2} \left( x
  \right) \dd x + O \left( \frac{1}{M^2 \Delta} \right) . \label{Fth}
\end{equation}
It is instructive to rewrite this
expression as part of the Cram\'{e}r-Rao inequality \cite{Giovanetti04} for the minimum average error on an unbiased estimate $\theta_{\text{est}}$ of a true phase $\theta$:
\begin{equation}
 \left\langle (\theta_{\text{est}} - \theta)^2 \right\rangle \geqslant \frac{1}{F_{\theta}} \approx \Delta + \frac{1}{j^2} \int \phi'^{\: 2} \left( x \right)
  \dd x. \tag{\ref{Fth}$'$} \label{Finv}
\end{equation}
Just how good is the approximation on the right side of  eqn.(\ref{Finv})? Consider a probe state having a
Gaussian profile $\phi_m \propto \exp (- m^2 / 2 w^2)$ with half-width $w \ll j$ (also known as `minimum uncertainty states' in the literature \cite{MUS}). The QFI  can be
evaluated exactly by diagonalizing the resulting Gaussian density matrix via Mehler's formula \cite{knysh} to yield $F_\theta = \left( \Delta + \frac{1}{4 w^2} \right)^{- 1}$, indicating the approximation is exact for Gaussian-profile states \cite{fn1}. A good example is the spin-coherent state occurring inside a Mach-Zehnder interferometer when all the probe light enters just one port of the first beamsplitter. Between the beam-splitters, these states have a Gaussian profile with half width $w = \sqrt{N}/2$, which gives $1 / F_{\theta} = \Delta + 1/ N$. Note that the latter statistical contribution to precision scales as shot noise. Obviously, better performance is possible for states with a wider distribution, but non-zero amplitude at the boundary $x = \pm 1$ will undermine precision. Otherwise the $w \sim \infty$ Gaussian-profile state, a `phase' state \cite{vourdas}, would be optimal. (Any discontinuity in $\phi$ at the boundary causes a spike in $\phi'$ in eq.\eqref{Fth} that in turn reduces the QFI.) So, not just the width, but the overall \emph{shape} of the profile is critical to reaching optimum precision, as we now discover.

Found by extremizing the leading contributions to the QFI functional in eqn.(\ref{Fth}), the probe state minimizing the phase error with support on the interval
$x \in \left[ - 1 ; 1 \right]$ is the Cosine function spanning half a period:
\begin{equation} \left| \psi_\text{opt} \right\rangle = \frac{1}{\sqrt{ j + 1/2}} \sum_{m =
   - j}^j \cos \frac{\pi m}{2 j + 1} \left| m \right\rangle \: ,\label{optimal}
   \end{equation}
yielding $1 / F_{\theta} \approx \Delta + \pi^2 / N^2$, the latter `non-classical' statistical contribution now obeying a `Heisenberg-like' quadratic scaling. (We call this contribution non-classical rather than quantum because our whole analysis is intrinsically quantum.) This partially entangled probe state is optimal even when dephasing noise is large, and its structure is largely independent of $\Delta$ in the $M \gg 1$ limit. In fact, the same optimal probe state is recovered by minimizing phase measurement error  $\langle \delta \theta^2 \rangle$ directly in the absence of dephasing,  see Ref.\cite{PeggSummy}, also producing $\langle \delta \theta^2 \rangle \approx  \pi^2 / N^2$. It is a subtle point that minimization of phase measurement error may \emph{not} correspond directly to optimization of precision, i.e. maximization of $F_\theta$. This is because, for non-Gaussian distributed probe states, the inequality $\langle \delta \theta^2 \rangle \geq \langle (\theta_{\text{est}} - \theta)^2 \rangle$ is not tight; an efficient estimator $\theta_{\text{est}}$ based on many data points may perform better on average than the measurement error for a single shot phase measurement (equivalent to the error on the sample mean). Consequently, optimizing QFI and phase variance can result in different optimal states, e.g. NOON state and Cosine state, respectively, for zero-dephasing case \cite{GaussOpt}. We shall return to this point later. 

The performance of the asymptotically optimal state is given quantitative comparison with other states proposed in the literature in FIG.2 for $j=100$ across a wide range of diffusion strengths. 

\begin{figure}
\includegraphics[width=3.5in]{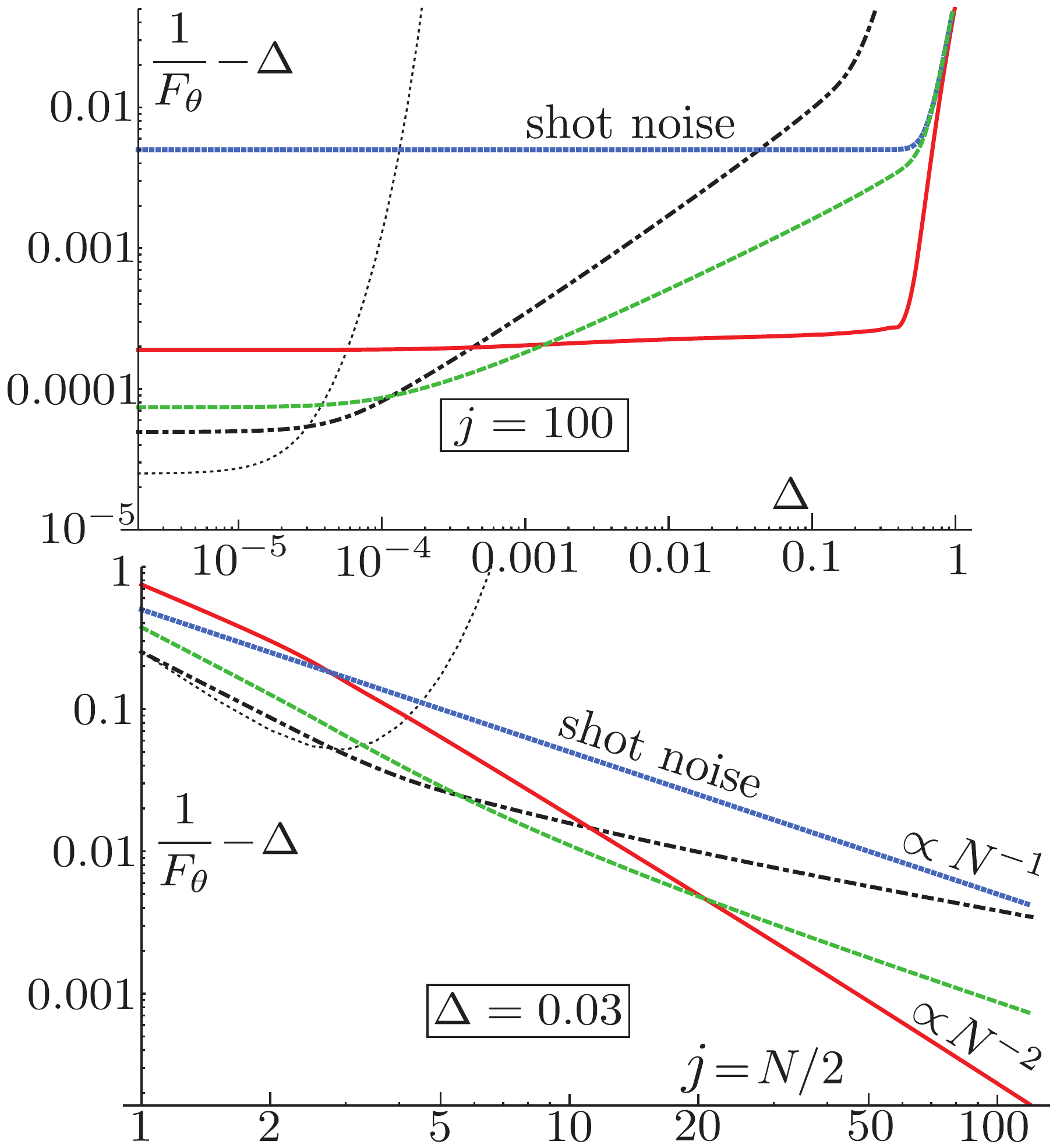}
\caption{\label{bars}
\small{Logarithmic plots of the `non-classical' contribution to overall minimum phase error: $1/F_{\theta} - \Delta$,  (i.e. after subtracting the purely classical phase noise), for fixed $j=100$ or $N=200$ particles (upper graph), and for fixed diffusion strength $\Delta =0.03$ (lower graph). The performance of the spin-coherent state (blue) with $\phi_m =  d^{j}_{m,j}(\pi/2)$  \cite{wigner} corresponds exactly to the shot noise limit for $\Delta \ll 1$. In the presented case, the asymptotically-optimal states (red curves) of eqn.\eqref{optimal} provide the minimum non-classical contribution to error $\approx \pi^2 / N^2$ across three orders of magnitude in the dephasing $\Delta$, although for very small dephasing $\sim 10^{-5}$ NOON states are optimal (faint dotted grey curves), as expected \cite{NOON}. Other states considered previously with intermediate performance are phase states \cite{vourdas} (green) with $\phi_m = 1 / \sqrt{2j+1}$ and Holland-Burnett states (black chained line) $\phi_m =  d^{j}_{m,0}(\pi/2)$ \cite{Holland93}. Notice the `sudden death'' of precision as $\Delta \gtrsim 1$, in which case the phase noise has become so great as to diagonalize the density matrix, making every input states useless in this limit. Note also the dominance of the optimal Cosine state of eqn.\eqref{optimal} for $j > 20$ suggesting that asymptotic behaviour is  apparent at moderately small particle number.}}
\end{figure}

\emph{Combining Errors and Optimal Measurements}: To make our results more intuitive, remember that phase diffusion is the addition of a classical random phase 
$\zeta$ to the interferometric phase $\theta$.
The leading order expression
eqn.(\ref{Finv}) is explicit in separating the total estimation error into that from $\left\langle \delta \zeta^2
\right\rangle = \Delta$ and the non-classical statistical uncertainty of estimating the total 
phase $ \theta + \zeta$ for a pure probe state using an optimal measurement (QFI assumes this implicitly). The foregoing discussion is equivalent to the realization that the dephased density matrix with damped off-diagonal elements is actually a Gaussian distributed mixture of pure probe states $ | \psi \rangle$, each shifted by a different phase $\theta + \zeta$:
\begin{equation}
\rho = \int_{-\infty}^{\infty} e^{-i J_z (\theta + \zeta)}  | \psi \rangle \langle \psi | e^{+i J_z (\theta + \zeta)}  | \frac{e^{-\zeta^2/ 2 \Delta}}{\sqrt{2 \pi \Delta}}     d \zeta
\end{equation}
If we imagine choosing one of these pure states from the mixture, let's make a `canonical' phase measurement \cite{phaseM} that projects the pure state onto phase states (see Ref. \cite{vourdas}): $\left| \theta_{\mu}
\right\rangle = \frac{1}{\sqrt{2 j + 1}} \sum_m \exp ( \! - \ii m \theta_{\mu})
\left| m \right\rangle$.
For symmetric probe distributions $\phi(x) = \phi(-x)$ such as those relevant to dephasing (itself a symmetric decoherence process), canonical phase measurements are globally optimal under unitary evolution by a phase shift such as $\theta + \zeta$, as was shown in \cite{CavesMilBraun}. Importantly, we can show that a mixture of such states will retain the \emph{same} optimal measurement in both $M \gg 1$ and $M \ll 1$ limits.

First, note that there are two classical probability distributions involved; that of the Gaussian-distributed random phase $p(\zeta)= e^{-\zeta^2/ 2 \Delta}/\sqrt{2 \pi \Delta}$ and the conditional distribution $p(\theta_{\mu}| \theta + \zeta) = |\langle \theta_\mu | \exp -  \ii (\theta + \zeta) J_z | \psi \rangle|^2$ associated with phase measurement result $\theta_{\mu}$ when phase evolution is $\theta + \zeta$ \cite{fn2}. The overall (covariant) probability distribution $\tilde{p}(\theta_\mu - \theta)$ is a \emph{convolution of both distributions}:
\begin{equation} \label{convP}
\tilde{p} \left( \theta_\mu - \theta \right)=\int_{2 \pi} p(\theta_{\mu}|\theta + \zeta) p(\zeta) (\delta \zeta) 
\end{equation}
Since $\theta_\mu$ and $\zeta$ are independent variables,  $\langle \delta \theta^2\rangle $ is simply the sum of their two variances (adding errors in quadrature). 

For the \emph{optimal} probe state evolved unitarily by phase $\theta$, a measure phase $\theta_{\mu}$ has a conditional probability $p(\theta_{\mu}|\theta) = p(\theta_{\mu} \! \! -\theta)  = | \langle \theta_{\mu} | \exp - \ii J_z \theta | \psi_\text{opt} \rangle |^2  $ that is non-Gaussian,
\begin{equation} \label{nonGaussian}
p(\theta_{\mu}|\theta) =\left( \frac{\sin \frac{\pi}{2 j + 1} \cos
 \{  \left( 2 j + 1 \right) (\theta_{\mu} \!   -\theta) \} }{\cos \frac{\pi}{2 j + 1} - \cos (\theta_{\mu} \! \ -\theta)}  \right)^2 ,
\end{equation}
 see FIG.3. This gives $ \langle (\theta_{\mu} \! \! -\theta)^2 \rangle  \approx \pi^2 / N^2$ in the absence of dephasing; adding it in quadrature to the classical diffusion variance $\Delta$  recovers the result for optimal precision $1 /F_\theta^{\text{opt}}$ to lowest order, with equivalence in the $M = \Delta j^2 \gg 1$ limit, as we had proposed.

\begin{figure}
\includegraphics[width=3.4in]{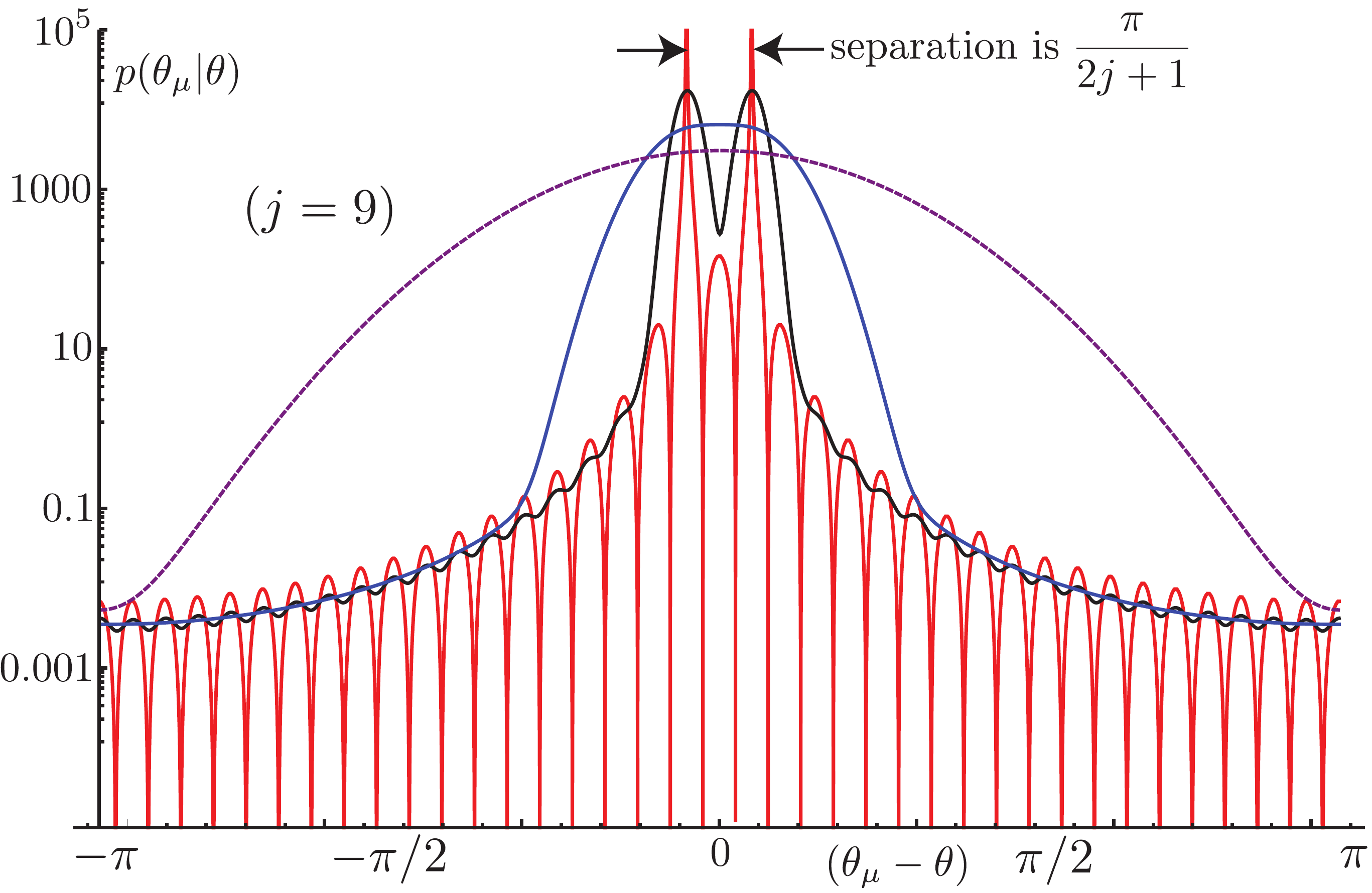}
\caption{\label{nonGauss}
\small{The non-Gaussian conditional probability distribution $p(\theta_\mu| \theta)$ from eqn.\eqref{nonGaussian} of a measured phase $ \theta_\mu$ when the true phase is $\theta$ for $j=9$. It depends only on the difference, $\theta_\mu -\theta$ (red curve).  On this logarithmic scale the Fisher information is equivalent to the negative of the average curvature for each distribution. The two central spikes of the optimal distribution contribute most to precision; the distribution is narrowly (and doubly) peaked near the true phase. This is the Fourier transform of the optimal Cosine state; if periodic boundary conditions were applied in the $x$ basis (rather than $\phi(x)=0$ for $|x|>1$) the above (red) distribution would consist only of two delta spikes.   The  `Gaussian-blurred' distributions $\tilde{p}$ from eqn.\eqref{convP}, recovered by convolving with classical phase noise of strength $\Delta \in \{0.003,0.03,0.3\}$, are plotted in black, blue and purple, respectively. A sort of Rayleigh criterion defines the threshold of Gaussianity: when the noise Gaussian has width greater than the distance between the two central spikes, i.e. $2 \sqrt{\Delta} >\pi / (2j+ 1)$, they can no longer be resolved; the convolved $\tilde{p}(\theta_\mu - \theta)$ becomes approximately Gaussian itself. Consequently, in the large mass $M = \Delta j^2 \gg 1$ (predominantly Gaussian) limit, the variance $ \langle (\theta_\mu - \theta )^2\rangle \sim \Delta + \pi^2/N^2$ of the phase distribution $\tilde{p}$ will be asymptotically equal to $1/F_\theta$ from eqn.\eqref{Finv}; a standard result for Gaussian statistics.  A corollary in this Gaussian limit is that the sample mean of the canonical phase measurements becomes the best unbiased estimator $\theta_{\text{est}}$.}}
\end{figure}

For clarification, let us proceed by writing an explicit chain of inequalities for optimal precision \emph{valid for all} $N \in [1,\infty]$:
\begin{equation}
  \frac{\Delta + \pi^2 / N^2}{\left[ 1 - 2 \pi \tilde{p} \left( \pi \right)
  \right]^2} = \langle \delta \theta^2\rangle   \geqslant  \left\langle (\theta_{\text{est}} - \theta)^2 \right\rangle \geqslant \frac{1}{F_{\theta}^{ \text{opt}}} \geqslant \Delta + 1 / N^2, \label{bounds}
\end{equation}
where the first equality merely states that for phase measurements the overall error is the sum of the classical noise and the non-classical measurement noise. The next inequality expresses the fact that the error on a single measurement $\theta_\mu$ is an upper bound to the error on the best unbiased estimate $\theta_{\text{est}}$ of the underlying interferometric phase after several data points have been collected.  To modify the Cram\'{e}r-Rao bound  with the denominator on the left hand side takes phase
periodicity into account as discussed in Ref.\cite{UncertPhase}. Here $\tilde{p}$ is the convolved distribution of  eqn.\eqref{convP}. Under ordinary circumstances, when $\Delta \ll 1$, the correction in the
denominator is $O \left( \Delta / j^2 \right)$ and can be neglected. In the
region $\Delta \gtrsim 1$ the denominator scales as $\ee^{- \Delta}$
 resulting in exponential rise in error. In FIG.2 this ``sudden death" of precision is indicated for a number of probe states in the large $\Delta$ limit.  
 
The final relation of \eqref{bounds} on the right side is the Fisher
information inequality for the sum of two independent random variables $F_{x +
y}^{- 1} \geqslant F_x^{- 1} + F_y^{- 1}$. This lower bound on error has
recently appeared in Ref. \cite{Escher12}. For fixed $\Delta$, by increasing $N$ the upper
bound $\Delta + \pi^2 / N^2$ on minimum error is saturated asymptotically (large `mass' $M$ limit) while the lower bound
$\Delta + 1 / N^2$ is appropriate in the small ``mass" limit $N \ll 1 / \sqrt{\Delta}$. 

Now it becomes clearer why minimization of $\langle \delta \theta^2\rangle $ for fixed $\Delta$ corresponds to the maximum QFI for large mass and is optimized by the same probe state; then the upper bounds to $1/F_{\theta}^{ \text{opt}}$ in \eqref{bounds} all become equalities. For large mass the convolved probability distribution $\tilde{p}$ will always be very close to Gaussian no matter how unclassical the measurement noise, it is dominated by the broad Gaussian $p(\zeta)$. Then no exotic estimator can improve on the precision bound provided by the one-shot measurement error.  To contrast, it is not appropriate to optimize $\langle \delta \theta^2\rangle $ for $M \ll 1$  as it does not provide a tight bound to precision $ \left\langle (\theta_{\text{est}} - \theta)^2 \right\rangle$ for multiple data. A more efficient estimator $\theta_\text{est}$ can be employed that exploits the non-Gaussian statistics of the probe to improve on the one-shot phase measurement error.

\emph{Clustering and Shot Noise}:  If we consider the $N$ particles as our resource to be divided how we please, we can devise an
optimal strategy for their use in phase estimation. By splitting them into $\nu$  clusters, each containing 
$N / \nu$ (possibly entangled) particles, we subject the clusters one at a time to the phase evolution and noisy environment. Estimating the optimal partitioning
requires further analysis to address the case
where $M \sim 1$. However, armed with both lower and upper bounds of eqn.\eqref{bounds} (valid for all $M$),  and
performing optimizations over $\nu$, we can write
\begin{equation}
  \frac{2 \sqrt{\Delta}}{N} \lesssim \frac{1}{F_\mathrm{opt}} \lesssim \frac{2
  \pi \sqrt{\Delta}}{N} \hspace{2em} \left( \Delta \ll 1 \right) \: .
\end{equation}
(The maximum is found by differentiating the bounds for the total Fisher information summed over the clusters with respect to $\nu$.) 
This result unequivocally establishes a shot-noise-like scaling of the error under collective dephasing. This
should be compared with the expression $\Delta / N$ for the minimum mean squared
error in a setting where each of the entangled qubits undergoes phase
diffusion locally and independently \cite{firstpaper,Escher12,RDD12,diffother}. (In a sense, collective dephasing is more deleterious to precision.) Dissipation, another type of noise, also results in unavoidable
asymptotic shot-noise scaling of precision \cite{knysh,losstoo}. 

We cannot easily determine the optimal partitioning into clusters; but if we compare the exact QFI expressions for a spin $j=1$ system and two unentangled $j=1/2$ particles ($2 e^{-\Delta}$) there is a critical dephasing $\Delta_c$ beyond which the strategy of sending the $N$ particles one at a time into the noisy environment will always perform better than utilizing clusters of higher spin (even bipartite $j=1$ clusters). It emerges that $\Delta_c \approx 0.2512$, adding credibility to the argument that  the limit $\Delta \ll 1$ is the important one for collective dephasing. Also, tripartite ( $j = 3/2$) clusters outperform bipartite ones for $\Delta < 0.081$, and $4$-part ($j=2$) clusters bypass tripartite clusters for $\Delta < 0.041$.

\emph{Estimation of Dephasing}: Quantum Fisher information $F_\Delta$ for estimation of $\Delta$ itself may be
computed in a similar fashion to the calculation of $F_\theta$. (As argued in the introduction, sensitive measurement of noise levels may be relevant for structural health monitoring and other applications.) Solving for the symmetric logarithmic
derivative
\[ \frac{1}{2} \left\{ \rho, j^2 L_{\Delta} \right\} = - \frac{1}{2} \left[
   jX, \left[ jX, \rho \right] \right] \]
is certainly less straightforward; performed to 4th order to capture the leading
behavior
\begin{multline}
  L_{\Delta} \approx - \frac{1}{2} \ddot{H} + \frac{1}{2} \dot{H}^2 +
  \frac{1}{24} \left[ H, \left[ H, \ddot{H} \right] \right] 
   + \frac{1}{12} \left[ \dot{H}, \left[ \dot{H}, H \right] \right] \\
  -\frac{1}{24} \left[ H, \dot{H} \right]^2 - \frac{1}{24} \left[ H, \left[ H,
   \dot{H}^2 \right] \right],
\end{multline}
where overdots denote commutation with $X$: $\dot{H}
\equiv \left[ X, H \right]$ and $\ddot{H} = \left[ X, \left[ X, H \right]
\right]$.

The expansion of $H$ itself in terms of commutators of $T$ and $U$ need only be
done to 3rd order as all even orders vanish due to symmetric form of
eqn.(\ref{eee}). The Fisher information is given as the expectation value of $-
\frac{1}{2} \ddot{L}_{\Delta}$. Collecting all the terms we obtain
\begin{equation}
  \frac{1}{F_{\Delta}} \approx 2 \Delta^2 + \frac{4 \Delta}{j^2} \int \phi'^2
  \left( x \right) \dd x \gtrsim 2 \Delta^2 + \frac{4 \pi^2 \Delta}{N^2}
  \label{FD}
\end{equation}
with the Cosine state of eqn.\eqref{optimal} also being optimal for the
estimation of $\Delta$. We can view this in terms of classical error analysis, as follows:
estimation of interferometric phase $\theta$ and dephasing parameter $\Delta$ is finding the mean and
 variance of the Gaussian distribution. If we again employ canonical phase measurements $\theta_\mu$, given results of $\nu$ independent
measurements, the unbiased estimator of the variance: $\frac{1}{\nu - 1}
\sum_{\mu = 1}^{\nu} \left( \theta_\mu - \bar{\theta} \right)^2$, is
$\chi^2$-distributed with variance $\frac{2 \Delta^2}{\nu - 1}$, corresponding
to the first term of eqn.(\ref{FD}). Including measurement noise $\delta \theta_\mu  = \theta_\mu - \zeta -\theta$ due to non-zero overlap $\langle \theta_\mu | \exp i (\theta + \zeta) |\psi_{\text{opt}}\rangle$ when $\theta_\mu \neq (\theta + \zeta)$, an additional contribution
 $\frac{4 \left\langle \delta \theta_\mu^2 \right\rangle}{\left( \nu -
1 \right)^2} \left\langle \sum_{\mu = 1}^{\nu} \left( \theta_\mu - \bar{\theta}
\right)^2 \right\rangle = \frac{4 \Delta \left\langle \delta \theta_\mu^2
\right\rangle}{\nu - 1}$ may be associated with the second term in eqn.(\ref{FD}).
A third term, proportional to the kurtosis of the distribution in eqn.\eqref{nonGaussian} can be neglected in the limit $M \gg 1$. To lowest order the classical error resulting from  canonical phase measurements agrees with the quantum Fisher information bound; these measurements are optimal for diffusion estimation.

Generally, in quantum estimation of multiple parameters ultimate bounds are
unachievable \cite{multi} since probe states yielding best precision may differ for each parameter, although exceptions exist \cite{joint2}.
Moreover, different measurements may be required to achieve individual quantum Fisher
information bounds.  Earlier work explored multiparameter estimation under
photon loss when such optimal states for phase \cite{knysh} and loss \cite{lossopt} estimation
are different. In addition, quantum uncertainty relations may conspire to make measurements
saturating quantum Fisher information bounds incompatible for any probe state \cite{joint}.
No such problems beset the simultaneous estimation of the dephasing parameter jointly with
the phase as both the optimal probe state and asymptotically optimal measurement are
identical. 

\emph{Summary and Outlook}: We investigated phase evolution and dephasing in the limit of a large number of qubits $N = 2j \gg 1$. We introduced a novel operator formalism, where  the dephased quantum system is represented as a particle of mass  $M = \Delta j^2$ subject to an abstracted Hamiltonian $H$. This enabled us to formulate quantum precision as a nested series of commutators of $H$ with the phase shift operator $X$. The first non-trivial contribution to phase precision is from a Bohmian quantum potential.

By this new operator approach, we found optimal states and a new tight saturable bound on phase precision, emphasizing the complementary nature of this bound with those already indicated in the literature as corresponding to the large and small `mass'  limits. For fixed dephasing noise the large mass limit will always be recovered for increasing $N$.

Insight is gained by understanding that the influence of the dephasing noise is to Gaussian-blur the optimal quantum phase distribution until it approaches a sort of Rayleigh limit, $\sqrt{\Delta} \sim 1/j$ for the dominant features of the distribution. For dephasing strength much beyond this limit the overall distribution of phase error becomes `Gaussianified' and there can be no efficient estimator that performs better than the error associated with a one-shot canonical phase measurement; the optimal state becomes the one that minimizes phase variance. Both unitary phase and non-unitary diffusion parameters are simultaneously and optimally measurable in the asymptotic limit $N >>1$ by  canonical phase measurements. 

Above a critical dephasing $\Delta_c \approx 0.2512$ entangled ensembles of particles evolved in parallel exhibit worse performance than subjecting them one particle at a time in series to the phase evolution and noisy environment. Generally, for optimal entanglement-clustering the phase estimation error $\sqrt{\langle ( \theta_{\text{est}} - \theta )^2 \rangle}$ is shot-noise limited for all $N$; this limit may only be surpassed by a constant factor, independent of $N$ but dependent on cluster size. For collective dephasing the phase estimation error is proportional to $\Delta^{1/4}$, as compared with $\sqrt{\Delta}$ for local dephasing models. Using our operator formalism  we were able, for the first time, to find the leading behaviour of the quantum Fisher information for estimation of diffusion strength. The lowest order contributions to precision for both phase and diffusion correspond to terms from classical error propagation.  

Our analysis is for a fixed system dimension, e.g. spins or flux qubits,  but remains valid for two-mode continuous-variable states of light, where the dimensionality is not fixed but rather expectation values like $\langle N \rangle$ are constrained \cite{fn3}. Future work might explore the evolution and structural bifurcations in the optimal state that occur as dephasing increases from the small to large mass limit; from a discrete 2-element NOON state to the smooth Cosine-profile. Another goal is to determine the best clustering of resources as a function of dephasing. These considerations, along with the results of this paper, point towards strategies for optimal design of next-generation real-world quantum sensors.

\appendix 
\section{Calculation of Quantum Fisher Information for Phase Estimation}
\label{append}
As a first step, we would like to solve
\begin{equation} \label{defSLD2}
  \frac{1}{2} \left\{ \rho, jL_{\theta} \right\} =  - \ii \left[ jX, \rho \right],
\end{equation}
for the symmetric logarithmic derivative $  L_{\theta}$ operator.
We write $\rho =
\ee^{- H}$, multiply eqn.(\ref{defSLD}) by $\ee^{H / 2}$ from the left
\emph{and} from the right and use the representation of $\ee^A B
\ee^{- A}$ as
\begin{equation} \exp \left( \left[ A, \bullet \right] \right) B \equiv B + \left[ A, B
   \right] + \frac{1}{2!} \left[ A, \left[ A, B \right] \right] + \cdots \end{equation}
with $\left[ A, \bullet \right] \equiv \text{ad}_A$ representing the adjoint
endomorphism of the corresponding Lie algebra. With the aid of this
identity, eqn.(\ref{defSLD2}) may be rewritten as $\cosh \left( \left[ \frac{H}{2},
\bullet \right] \right) L_{\theta} = - 2 \ii \sinh \left( \left[ \frac{H}{2},
\bullet \right] \right) X$ and finally,
\begin{equation}
  L_{\theta} \! =  \!- 2 \ii \tanh  (  \left[ H, \bullet \right] /2) X\!  \approx \!  - \ii [ H, X ] + \frac{\ii}{12} [ H, [ H, [ H, X ]]]  
\end{equation}
with successive terms corresponding to the Taylor expansion of hyperbolic tangent. Now we  can be express the QFI in terms of this operator, as follows:
\begin{equation}
  F_{\theta} / j^2 = \text{Tr} \left( \rho L_{\theta}^2 \right) = \text{Tr} 
  \left( - \ii \left[ X, \rho \right] L_{\theta} \right) = \left\langle
  \ii \left[ X, L_{\theta} \right] \right\rangle,
\end{equation}
where angle brackets denote trace with the density matrix. We will retain only the leading and next-to-leading orders in the BCH expansion of $H \approx H_0 +H_1$ from eqn.\eqref{BCH}, i.e. $H_{0} = T + U$, and 
\begin{equation} H_1 \! = \!  \frac{1}{12} [T, [T, U
 ]]- \frac{1}{24} [ U, [ T, U ] ] \! = \! - \frac{\left\{ P, \left\{ P, U'' \right\} \right\}}{48 M^2} + \frac{U'^2}{24 M},
\end{equation} 
with curly braces denoting
anticommutators and derivatives with respect to $x$ indicated by primes. The leading
contribution $\left\langle \left[ X, \left[ H_0, X \right. \right]
\right\rangle$ yields a constant $F_{\theta}^{\left( 0 \right)} / j^2 = 1 /
M$. Next order corrections to $F_{\theta} / j^2$ \ due to 3rd order terms in
eqn.(\ref{BCH}) and eqn.(\ref{SLD}) are given as the expectation value of the Bohmian quantum potential:
\begin{equation} 
\left[X, \left[ H_1, X \right] \right] - \frac{1}{12} \left[ X, \left[
   H_0, \left[ H_0, \left[ H_0, X \right] \right] \right] \right] = -
   \frac{U''}{4 M^2}, \end{equation}
evaluated by integrating it with weight $\phi^2 \left( x \right)$.
Substituting $U = - \ln \phi^2 \left( x \right)$ and proceeding to integrate by parts, remembering the boundary condition $\phi = 0$ for $|x|>1$ gives the result presented in eqn.\eqref{Fth}:
\begin{equation}
  F_{\theta} = \frac{1}{\Delta} - \frac{1}{M \Delta} \int \phi'^{ \: 2} \left( x
  \right) \dd x + O \left( \frac{1}{M^2 \Delta} \right) . 
  \end{equation}
  
 \pagebreak


\begin{thebibliography}{99}
\bibitem{fn1}
In this particular case the approximate formula gives the correct result for all
$\Delta \ll 1$ and $j \gg 1$; the parameter $M = j^2 \Delta$ need not
necessarily be large.
\bibitem{GaussOpt} Boundary conditions are just as important in determining the optimal state. If no boundary conditions are applied at all, optimizing phase variance actually predicts the Gaussian-profile state is optimal, as shown in \cite{CavesMilBraun}.
\bibitem{fn2}
This distribution is generally not a delta function because the canonical phase measurements form a non-orthogonal overcomplete set, and because the probe state is not generally a phase state either.
\bibitem{fn3}
In most of these systems the measure observables ultimately involve detection of particle number, projecting into a space of a specific $N$. (Even measurement of field quadratures involve photon number measurement of ancillary local oscillators.) The Fisher information over uncorrelated particle number spaces is simply the ensemble average of that in each particle number space.
\end{thebibliography}
\end{document}